\input epsf
\documentstyle[12pt]{article}
\newcommand{\be}{\begin{equation}}
\newcommand{\ee}{\end{equation}}
\newcommand{\beq}{\begin{eqnarray}}
\newcommand{\eeq}{\end{eqnarray}}

\begin{document}

\thispagestyle{empty}
\rightline{NSF-ITP-94-34}
\rightline{hep-th/9405084}

\vskip 2.5 cm

\begin{center}
{\large\bf  Conformally Invariant Boundary Conditions \break
for Dilaton Gravity} \break

\vskip 1.0 cm

{\bf Andrew Strominger} \footnote{Electronic address:
andy@denali.physics.ucsb.edu}

{\sl Department of Physics \break
University of California \break
Santa Barbara, CA 93106-9530} \break

and {\bf L{\'a}rus Thorlacius} \footnote{Electronic address:
larus@nsfitp.itp.ucsb.edu}

{\sl Institute for Theoretical Physics  \break
University of California  \break
Santa Barbara, CA 93106-4030} \break

\end{center}

\vskip 1.0 cm

\begin{quote}

Quantum mechanical boundary conditions along a timelike line,
corresponding to the origin in radial coordinates, in
two-dimensional dilaton gravity coupled to $N$ matter fields,
are considered. Conformal invariance and vacuum stability
severely constrain the possibilities. The simplest choice found
corresponds to a nonlinear Liouville-type boundary interaction.
The scattering of low-energy matter off the boundary can be
computed perturbatively. It is found that weak incident pulses
induce damped oscillations at the boundary while large incident
pulses produce black holes.  The response of the boundary to such
pulses is semi-classically characterized by a second order, nonlinear
ordinary differential equation which is analyzed numerically.
\end{quote}

\newpage

\section{Introduction}

Two-dimensional dilaton gravity coupled to conformal matter provides
a simple framework for studying the puzzles of black hole formation
and evaporation. In some models, a fairly complete understanding of
the leading term in a $1/N$ expansion (where $N$ is the number of
matter fields) of the quantum theory has been obtained
\cite{cghs}-\cite{revs}.  Below a certain energy threshold, black holes are
not formed (at this level of approximation).  Above the threshold,
black holes form and evaporate, and information is irretrievably lost
into spacelike singularities.

The relevance of these results to the black hole information puzzle can
be legitimately questioned on at least two grounds. First, despite the
fact that it contains black holes and Hawking evaporation, the theory
may not faithfully model the four-dimensional phenomena. Physics in
two dimensions is certainly rife with peculiarities.
Second, it is not understood -- even
in principle -- how to compute the subleading corrections to
the leading large-$N$ behavior. Even a formal definition of the exact
quantum theory has not been given for these models. Thus it cannot be
claimed that any fully self-consistent model with information loss exists.
Clearly it is important to establish whether or not this is the case.

One of the main issues at stake here is the nature of the
boundary conditions imposed along a timelike line which lead to the
reflection of  below-threshold incoming energy \cite{rst}. Such a
boundary is required in order that the two-dimensional theory faithfully
model the desired four-dimensional physics.  The two dimensions
correspond to time and the half-line $r>0$. The boundary corresponds
to the origin of the four-dimensional, spherically symmetric
spacetime.  A below-threshold pulse which is reflected off of the boundary
in the two-dimensional model corresponds to a low-energy four-dimensional
S-wave which passes through the origin without gravitationally
collapsing.

The boundary conditions are highly constrained by the
following consistency conditions:

\begin{enumerate}
\item{ {\em Conformal Invariance.\/} General covariance and energy
conservation require that the boundary conditions are conformally
invariant.}
\item{ {\em Vacuum Compatibility.\/} The linear dilaton vacuum --
or a close cousin -- must be compatible with the boundary conditions.
If the nature of the vacuum is greatly altered, the model can not be used
to study black hole physics.}
\item{ {\em Vacuum Stability.\/} The boundary conditions should insure that
that the vacuum is stable under small perturbations.}
\end{enumerate}

In this paper we build on earlier work
\cite{rst,verl,verdan,das} and analyze the problem
semi-classically. A solution is found only when $N>24$ (but is not
necessarily ``large").
In the large-$N$ limit it agrees with that found previously
by Chung and Verlinde \cite{verdan}.
The solution involves an exponential Liouville-type boundary
interaction. There is one free parameter, $Y_0$ which governs the strength
of the boundary interaction, but this parameter can not be dialed to turn
off the interaction without encountering instabilities.
The result of throwing low-energy pulses at the boundary can be computed
perturbatively.  For a range of $Y_0$ and $N>24$, the incoming pulse
excites damped oscillations of the boundary\footnote{Depending on the
gauge condition, either the boundary curve itself or fields at the
boundary oscillate.}. For values of $Y_0$ outside this range, incoming
pulses excite exponentially growing oscillations and vacuum instability.
When $N<24$, there is no stable range of $Y_0$.  (Although alternate
boundary conditions, which give stable dynamics for $N<24$, may well exist.)
The interesting special case $N=24$ will be described elsewhere \cite{stt}
and has been previously discussed from a somewhat different perspective
in \cite{verl,das}.

For the string theorists in the audience, we note that the question of
consistent boundary conditions can be viewed as a problem in open string
theory. In that language, our boundary conditions correspond to an
open string tachyon condensate.

All the results of this paper are based on a semi-classical
approximation. Operationally it is clear how to implement this
approximation. It is much less clear when the approximation is reliable.
It is generally justified only at
large $N$, so our strongest results pertain to the large-$N$ limit. At
finite $N$ the approximation is still justified for the computation of
certain quantities in coherent semi-classical states, and
may generally provide insight in to the structure of the theory.
However  we wish to
caution the reader that its reliability is much more limited at finite $N$.
While we do believe that exactly conformally invariant boundary
conditions of the type we describe exist at finite $N$, the methods of this
paper are insufficient to establish that.  A fully quantum treatment may be
possible when $N=24$, where substantial simplifications occur.

In Section~\ref{secii} we establish our notation
and review the transformation
of the bulk theory of two dimensional dilaton gravity to a
soluble conformal field theory \cite{abc,dea,qtdg}.
We also show that black holes in these theories
evaporate at a rate proportional to $N$ -- even when $N<24$ -- as desired.
(The literature contains conflicting claims on this point.)
In Section~\ref{seciii} our boundary conditions are presented and analyzed,
and in Section~\ref{largen} we consider their large-$N$ limit.
Section~\ref{seciv} analyzes the short-distance limit
of the theory, in which
the bulk cosmological constant can be ignored. This limit is of special
interest because the full theory, including the exponential boundary
interaction, is semi-classically (and possibly exactly) soluble.

The present work on low-energy scattering is a prerequisite to, but still a
long way from, a fully consistent quantum description of
black hole formation and evaporation as approximately described
at large $N$. In particular we do not address the issue of how the
black hole disappears after shrinking to the Planck size. This black
hole endpoint is in general out of causal contact
with the point where the collapsing matter arrives at the origin
(as discussed in Section~\ref{seciii}) and so cannot be affected by any
boundary conditions we impose there. We leave this vexing issue for future
work \cite{asbs}.

\section{The Bulk Theory}
\label{secii}

In this section we describe the bulk conformal
field theory of two-dimensional dilaton gravity for arbitrary $N$,
deferring the issue of boundary conditions to the next section.
Our aim is mainly to refresh the reader's memory and fix
conventions, but we also present some new material on the rate of Hawking
evaporation of large mass black holes.
The reader is referred to \cite{revs} for more thorough reviews of
the subject.

\subsection{Quantization}

Classical dilaton gravity is described by the action
\be
S_{classical}=\frac{1}{2\pi}\int d^2\sigma \sqrt{-g}e^{-2\phi}
\left[R+4(\nabla\phi)^2 +4\lambda^2- \frac{1}{2} \sum^N_{i=1} (\nabla
f_i)^2 \right].
\label{clct}
\ee
In conformal gauge,
\be
ds^2=-e^{2\rho}d\sigma^+d\sigma^-, \label{cggm}
\ee
where $\sigma^\pm=\sigma^0 \pm \sigma^1$, the action becomes
\be
S_{classical}=\frac{1}{\pi} \int d^2\sigma [-2\partial_+ e^{-2\phi}
\partial_-(\rho-\phi)+
\lambda^2 e^{2(\rho-\phi)}
+\frac{1}{2} \sum^N_{i=1} \partial_+f_i\partial_-f_i]\,.
\label{cgct}
\ee
At the one-loop level, the action (\ref{cgct}) acquires a well-known
correction
\be
-\frac{N}{12\pi} \int d^2\sigma \partial_+\rho\partial_-\rho \label{rrmt}
\ee
from the functional measure for the matter
fields. The accompanying correction to the stress
tensor will be given below.
A similar correction to (\ref{clct}) arises from the
$\rho$, $\phi$ and ghost measures \cite{strgh},
\be
\frac{2}{\pi} \int d^2\sigma
\partial_+(\rho-\phi)\partial_-(\rho-\phi).
\label{rrgh}
\ee
The combination $\rho-\phi$ (rather than $\rho$) appears in (\ref{rrgh})
because the natural metric for the $\rho,\phi$ and ghost fields is
$ds^2=-e^{2\rho-2\phi}d\sigma^+d\sigma^-$, rather than (\ref{cggm}).
This is the metric which appears in the kinetic term for $\rho$ and
$\phi$ in the classical action (\ref{cgct}).
We will also see that this choice of measure is
required in order that black holes
Hawking evaporate at a rate proportional to $N$, rather than $N-24$.

The choice (\ref{cggm}) of conformal gauge leaves unfixed a group of
residual coordinate transformations which is isomorphic to the conformal
group. Thus in order to maintain coordinate invariance of the quantum
theory, one must ensure that conformal invariance of (\ref{cgct}) --
regarded as a theory of the $N+2$ bosons $f_i,\rho$, and $\phi$ --
is preserved by the quantization procedure. The sum of the classical
action (\ref{cgct}) plus the one loop corrections (\ref{rrmt})
and (\ref{rrgh}) is not conformally invariant to all orders in the loop
expansion parameter $e^{2\phi}$ (although it is conformally invariant to
leading order in the $1/N$ expansion).
To remedy this we must add counterterms at each order. Conformal invariance
does not uniquely fix these counterterms. Different choices lead to
inequivalent but in general qualitatively similar theories.
It behooves us to choose the counterterms so as to simplify theory as
much as possible. As discussed in \cite{dea,abc,qtdg}, a
judicious choice leads to a soluble conformal field theory.
Let\footnote{A separate treatment will be given for $N=24$
$(\gamma=0)$ in \cite{stt}.}
\beq
\gamma&=&\frac{N-24}{12} \,, \nonumber\\
Y&=&-\frac{1}{\gamma} \int d\phi\sqrt{4e^{-4\phi}
-4(\gamma+2)e^{-2\phi}+2(\gamma+2)} \,,
\nonumber\\
X&=&\rho + \frac{e^{-2\phi}+2\phi}{\gamma} \,.
 \label{xdfn}
\eeq
The action is then given by
\be
S=\frac{1}{\pi} \int d^2\sigma \left[-\gamma\partial_+X\partial_-X
+\gamma\partial_+Y\partial_-Y +\lambda^2e^{2(X-Y)}
+\frac{1}{2} \sum^N_{i=1} \partial_+f_i\partial_-f_i\right] \,.
 \label{xctn}
\ee
Substitution of (\ref{xdfn}) into this action leads back to the original
action (\ref{cgct}) (corrected by (\ref{rrmt}) and (\ref{rrgh})),
plus additional potential terms which are subleading in the loop expansion
parameter $e^{2\phi}$. These corrections ensure exact conformal invariance.

The gravitational and matter parts of the stress tensor are
given by\footnote{Our normalization follows prevalent
dilaton gravity conventions, which
unfortunately differ from prevalent conformal field theory
conventions.}
\be
T^g_{++}=
\gamma\left(
  \partial_+Y \partial_+Y
- \partial_+X \partial_+X
+ \partial^2_+X \right) \,,\qquad
T^f_{++}=\frac{1}{2}\sum^N_{i=1}\partial_+f\partial_+f \,,
 \label{xyst}
\ee
along with a similar expression for $T^g_{--}$, $T^f_{--}$.
Together they generate a $c=26$ Virasoro algebra.

The theory defined by  (\ref{xctn}) is almost as simple as a free field
theory. Indeed, in Section~\ref{seciv} we show how it can be transformed
into one.  The null combination $(X-Y)$
which appears in the exponential  in (\ref{xctn}) does
not have a singular  operator product with itself, and a
superselection rule prevents the exponential interaction term
from generating corrections to the free OPE's.
For these reasons the theory is substantially simpler than the
Liouville theory, which it superficially resembles. For
example, the quantum effective action can be (formally)
exactly computed and is the same as the original action
(\ref{xctn}) \cite{rst}. However, these simplifications evaporate
to a large extent when the boundary
is introduced, as we shall see in Section~\ref{seciii}.

It is instructive to consider the semi-classical solutions to this
theory.  By this we mean solutions of (\ref{xctn}), rather than
of the original classical action (\ref{clct}). Thus our semi-classical
solutions will contain terms of arbitrary order in $e^{2\phi}$, and
know about Hawking radiation and black hole evaporation.
The semi-classical equations of motion have a two parameter family of
static solutions labeled by the constants $\mu$ and $F$
\beq
\gamma X &=&  e^{2\lambda\sigma} +
(F+\frac{\gamma}{2})\lambda\sigma
+\frac{\mu}{\lambda}, \nonumber\\
Y &=& X - \lambda \sigma \,,
 \label{ssh}
\eeq
where $\sigma=\frac{1}{2}(\sigma^+-\sigma^-)$.  We have expressed the
solutions here in ``sigma gauge'', generally defined by the condition
$Y= X - \lambda \sigma$, in which the static field configurations are
independent of the timelike coordinate. For all these solutions
$\phi\to-\lambda\sigma$ and $\rho\to0$ asymptotically.
The vacuum corresponds
to $\mu=0$ and $F=-1$ \cite{qtdg}:
\beq
\gamma X &=& e^{2\lambda\sigma} +
(\frac{\gamma}{2}-1)\lambda\sigma , \nonumber\\
Y &=& X - \lambda\sigma \,.
 \label{vcmm}
\eeq
The gravitational stress tensor takes the value
$T^g_{++} = \frac{\lambda^2}{2}$ in the vacuum configuration (\ref{vcmm}).
The non-vanishing right hand side of this equation
cancels (in sigma coordinates)
against the ghost stress tensor, $T^{gh}_{++}=2(\partial_+(\rho-\phi))^2$,
which is included in $T^g_{++}$ in
the left hand side of the equation. Equivalently,
the ghost vacuum is annihilated by
annihilation operators defined in the (Kruskal) coordinates in which
$\rho=\phi$ asymptotically\footnote{As such it is closely related to
the familiar $-1$ shift of $L_0$ in string theory. We thank S. Trivedi for
pointing this out.}. Although this seems the most natural
choice, a priori, other choices of the ghost vacuum
state might be considered. However, we shall see in the next subsection
that black holes evaporate at a rate proportional to $N$ only if
$F=-1$.

\subsection{Black Hole Formation and Evaporation}

Consider matter incident on the vacuum (\ref{vcmm}) from ${\cal I}^-$
characterized by some given energy profile $T^f_{++}(\sigma^+)$.
The semi-classical gravitational field is a solution of the constraints
\beq
T^g_{++} &=& -T^f_{++}-\frac{F\lambda^2}{2} , \nonumber\\
T^g_{--} &=& -\frac{F\lambda^2}{2}. \label{cstr}
\eeq
This is asymptotic to the vacuum (\ref{vcmm}) only when $F=-1$, but for
the moment we consider arbitrary values of $F$ in order to see why
$F=-1$ is indeed the correct choice.  The solution of these equations is
\beq
\gamma X &=&
e^{2\lambda\sigma}
-\frac{1}{\lambda}e^{\lambda\sigma^+} P_+(\sigma^+)
+\frac{1}{\lambda}M(\sigma^+)
+(F+\frac{\gamma}{2})\lambda\sigma \,, \nonumber\\
Y &=& X - \lambda\sigma \,,
 \label{xysl}
\eeq
where
\beq
M(\sigma^+) &=& \int_{-\infty}^{\sigma^+} T^f_{++}(v^+) dv^+ \,,
\nonumber\\
P_+(\sigma^+) &=& \int_{-\infty}^{\sigma^+} e^{-\lambda v^+}
T^f_{++}(v^+) dv^+.
 \label{mpdf}
\eeq
The Bondi mass measured on ${\cal I}^+$ is given by
\be
m(y^-)=2e^{\lambda(y^+-y^-)}
(\lambda\,\delta\rho+\partial_+\delta\phi-\partial_-\delta\phi) \,.
 \label{mbnd}
\ee
This expression is evaluated in asymptotically inertial coordinates
$y^\pm$ in which $\rho\to0$, and $\delta\phi$ and $\delta\rho$ are the
deviations of $\phi$ and $\rho$ from their vacuum values.
The inertial coordinates $y^\pm$ at ${\cal I}^+$ are
\beq
y^+ &=& \sigma^+,\nonumber\\
y^- &=& \sigma^--\frac{1}{\lambda}
\ln(1-\frac{1}{\lambda}\bar{P}_+e^{\lambda\sigma^-})\,,
 \label{yztf}
\eeq
where $\bar{P}_+ \equiv P_+(\infty)$.  To leading order in
$e^{-\lambda\sigma^+}$, one finds
\be
\delta\phi=-\frac{\delta Y}{2Y} = \delta\rho \,,
 \label{dlts}
\ee
where
\be
\gamma \delta Y=\frac{\bar{M}}{\lambda}
-\frac{1}{4}(\gamma-2F)
\ln(1+\frac{1}{\lambda}\bar{P}_+e^{\lambda y^-}) \,,
 \label{dlty}
\ee
with $\bar{M} \equiv M(\infty)$.  The Bondi mass is then
given by
\be
m=\bar{M}-\frac{\lambda}{2} (\frac{N}{24}-1-F)\left(
\ln  (1+\frac{1}{\lambda}\bar{P}_+e^{\lambda y^-})
+\frac{\bar{P}_+}{\bar{P}_+ +\lambda e^{-\lambda y^-}}\right)\,.
 \label{mfnl}
\ee
The rate of change of the mass is evidently proportional to $N$ if and
only if $F=-1$. We have stressed this point because of persistent
confusion in the literature.

At early retarded times $(y^-\to-\infty)$ the mass (\ref{mfnl}) decays
at the rate predicted by naive semi-classical reasoning which ignores
backreaction. At late times $(y^-\to+\infty)$, however, a disaster occurs:
the mass plummets to minus infinity. This is not so surprising because
$\mu$ in eq.~(\ref{ssh}) can be arbitrarily negative and the
system has no ground state.
The ``vacuum'' (\ref{vcmm}) is unstable under arbitrarily small
perturbations in the context of the bulk theory considered so far.
We shall see that this disaster is averted for
small $\bar{P}_+$ when appropriate boundary conditions are imposed.

\section{Boundary Conditions}
\label{seciii}

In the bulk theory of the preceding section,
the range of values taken by $X$
and $Y$ is unrestricted. For $\gamma>0$ this is unphysical because taking
$Y$ below a certain minimum value corresponds, in this case, to
complex values of the original dilaton field $\phi$ \cite{rst}.
When two-dimensional
dilaton gravity is derived by spherical reduction from four dimensions,
this corresponds to transverse two-spheres of negative area.
The bothersome negative energy configurations are also characterized by
large regions in which $Y$ is below this minimum.  For
$\gamma<0$ the field redefinition (\ref{xdfn}) is
non-degenerate for all values of $Y$ but the bulk theory
has nevertheless the same negative energy problem.

One can try to remedy these problems by simply
restricting the range of $Y$ so that the original fields $\rho$ and $\phi$
are real. However, the
resulting functional integral no longer defines a conventional quantum
field theory, and in particular does not correspond to any easily
identifiable conformal field theory.

An alternate procedure, which does lead to a conformal field theory, is to
impose boundary conditions along a timelike line
$\sigma=constant$ at or near the ``origin''.
Incoming fields will then be reflected at this line to
outgoing fields. In the physical region to the right of this line, the
semi-classical vacuum values of $X$ and $Y$ will
correspond to real values of $\rho$ and $\phi$.
Strong quantum fluctuations -- or large incoming pulses -- will still
produce low values of $Y$, but this does not preclude the perturbative
construction of a low-energy $S$-matrix for asymptotic observers.

One might hope to also effect a cure of the negative energy instability
of the $\gamma<0$ theory by imposing a boundary condition at some value
of $Y$, even if there is no natural minimum value in that case.  As we
shall see, however, the solutions exhibit a qualitatively different
behavior for $\gamma<0$ and we have not found any boundary conditions
which stabilize that theory.

\subsection{Conformal Invariance and Vacuum Compatibility}

It is imperative that the boundary conditions respect conformal invariance.
Otherwise the theory is not generally covariant
and cannot be regarded as a theory of gravity. Boundary
conformal invariance is equivalent to \cite{cardy}
\be
T_{++}(0,\tau)=T_{--}(0,\tau), \label{ttpm}
\ee
where $T$ is the total stress tensor for all fields. If either Dirichlet
or Neumann boundary conditions are imposed on the matter fields $f_i$:
\be
\partial_+f_i(0,\tau) \pm \partial_-f_i(0,\tau)=0, \label{dnff}
\ee
then the boundary condition (\ref{ttpm}) implies
\be
\partial^2_+X-\partial_+X\partial_+X +\partial_+Y\partial_+Y
= \partial_-^2X-\partial_-X\partial_-X +\partial_-Y\partial_-Y.
\label{xxyy}
\ee
This equation is {\it semi-classically} solved by either
\beq
\partial_+X - \partial_-X&=&\lambda F(Y)\,e^X, \nonumber\\
\partial_+Y - \partial_-Y&=&-\lambda \frac{\partial F}{\delta Y}\,e^X,
 \label{bcnn}
\eeq
where $F(Y)$ is an arbitrary function,\footnote{In the
language of string theory this corresponds to an open string tachyon,
$T=F(Y)e^{X}$, which satisfies the classical beta-function condition,
$\nabla\Phi\cdot\nabla T=8T$.} or
\beq
\partial_+X - \partial_-X&=&\lambda Ae^{X-Y_0},\nonumber\\
Y&=&Y_0,
 \label{bctw}
\eeq
where $A$ and $Y_0$ are constants.  In the following we will focus on
(\ref{bctw}). The imposition of these boundary conditions transforms the
relatively trivial bulk theory to a highly non-linear, interacting theory.

Compatibility of these boundary conditions with the vacuum solution
(\ref{vcmm}) constrains the parameters $A$ and $Y_0$. For a given
value of $A$, the $X$ boundary condition is satisfied for the vacuum
solution (\ref{vcmm}) only along the
timelike line
\beq
\lambda\sigma&=&\omega_0,\nonumber\\
\gamma A  &=& 2e^{\omega_0}
+ \frac{(\gamma-2)}{2}e^{-\omega_0} ,
\label{tmln}
\eeq
along which
\be
\gamma Y=e^{2\omega_0}-\frac{(\gamma+2)}{2}\omega_0.
\label{tmsn}
\ee
$A$ and $Y_0$ are therefore determined by the single free parameter
$\omega_0$\footnote{This constraint could be
relaxed by considering $\mu \neq 0$ solutions in (\ref{ssh}),
but we have not explored this possibility.}.
For a given value of $A$ and $\gamma >2$, there are two values of
$\omega_0$ consistent with (\ref{tmln}). However we shall see that
at most one is stable.
Note also that there is a non-zero minimum value of $A$,
\be
A_{min}= \frac{2\sqrt{\gamma-2}}{\gamma},
\label{amin}
\ee
for $\gamma >2$. Thus it is not in this case possible to analyze the
theory perturbatively in the strength of the boundary interaction.

The boundary conditions (\ref{bcnn}) and (\ref{bctw}) are
only semi-classical solutions of the operator condition (\ref{ttpm})
for conformal invariance. To verify that they imply the reflection
condition (\ref{ttpm}) on the stress tensor to leading order, $X$ and $Y$
are treated like $c$-number fields. The boundary conditions can
presumably be modified order by order in the loop expansion of (\ref{xctn})
in order to maintain (\ref{ttpm}).

Of course, the semi-classical approximation is not always reliable.
Corrections to the semi-classical approximation can be systematically
suppressed by taking $\gamma$ to be large, as will be discussed in
Section~\ref{largen}. However, even when $\gamma$
is not large, the semi-classical approximation can be good for certain
quantities or certain states. For example we expect the semi-classical
approximation to be good for calculating the radiation rate of a large
black hole for any value of $\gamma$.
We also expect that a necessary condition
for the semi-classical approximation
(\ref{bctw}) to the boundary conditions to be good is that the boundary
is in a weak-coupling, large-radius ({\it i.e.} large $Y$) region.
It follows immediately from (\ref{tmsn}) that it is
always possible to arrange that this is the case by adjusting the free
parameter $A$ in (\ref{bctw}) to be very large.

\subsection{Dynamical Boundary Curve}

A boundary condition imposed at $\sigma=constant$ restricts the left and
right conformal invariance to a diagonal subgroup. Separate left and right
invariance can be regained, however, at the price of allowing the boundary
to follow a general trajectory, described by the equation
$x_B^-(x^+)=x^-$. In this case constancy of $Y$
along the boundary implies
\be
\frac{1}{u}\partial_+Y+u\, \partial_-Y=0, \label{ybcg}
\ee
where
\be
u(x^+)\equiv
\left(\frac{\partial x_B^-}{\partial x^+}
\right)^{1/2},
 \label{udfn}
\ee
and $(\frac{1}{u},u)$ is the tangent vector to the boundary
curve. The Neumann or Dirichlet
conditions on the matter fields become
\be
\frac{1}{u}\partial_+f_i\pm u\,\partial_-f_i=0 \,.
 \label{fbcg}
\ee
The general form of the $X$ boundary conditions follows readily from the
observation that (\ref{bctw}) is equivalent
to the geometric condition that the extrinsic curvature of the boundary
curve in the metric $ds^2=-e^{2X}dx^+dx^-$ is constant.
For a general curve the first equation of (\ref{bctw}) becomes
\be
\frac{1}{u}\partial_+X- u\,\partial_-X
=\lambda Ae^{X-Y_0}+\frac{1}{u^2}\partial_+u.
 \label{xbcg}
\ee

The boundary conditions (\ref{ybcg}) and (\ref{xbcg}) can be used to
relate the components $T^g_{++}$ and $T^g_{--}$
of the gravitational stress tensor along the boundary.
Since (\ref{xbcg}) holds everywhere along the boundary, a new
identity can be obtained by acting on both sides with the operator
$\frac{1}{u}\partial_++u\partial_-$
which generates translations along the boundary. One finds
\be
\frac{1}{u^2}(\partial_+^2X-\partial_+X\partial_+X)
= u^2 (\partial^2_-X - \partial_-X\partial_-X)
- \frac{1}{u}\partial_+^2\left(\frac{1}{u}\right)\,.
 \label{xnml}
\ee
Taken together with the $Y$ boundary condition (\ref{ybcg}) this implies
that
\be
\frac{1}{u^2}T^g_{++}=u^2T^g_{--}-
\frac{\gamma}{u}\partial_+^2\left(\frac{1}{u}\right) \,.
 \label{tgpm}
\ee
The last ``Schwinger" term is well known in studies of moving mirrors
\cite{mirrs}. It vanishes in ``straight line" gauges for which $u$ is
constant. This Schwinger term
was omitted in the reflecting boundary conditions discussed
in \cite{rst}.  The boundary conditions of \cite{rst} can apparently not be
derived as the semi-classical limit of any conventional quantum mechanical
boundary conditions\footnote{This observation was made in collaboration
with S.~Trivedi.}.  They nevertheless
give rise to a stable evolution which
conserves semi-classical energy.

A differential equation for the boundary curve can be easily derived in
Kruskal gauge which is defined by the condition
\be
X(x^+,x^-)=Y(x^+,x^-) \,,
 \label{kcgn2}
\ee
where Kruskal and sigma coordinates are related by
\be
\lambda x^\pm=\pm e^{\pm \lambda \sigma^\pm}.
\label{krsig}
\ee
Adding (\ref{ybcg}) and (\ref{xbcg}) we then find
\be
\frac{2}{u} \partial_+ Y = \lambda A+\frac{1}{u^2}\partial_+ u \,.
 \label{wtfd2}
\ee
Multiplying by $u$ and acting on both sides with $(\partial_++u^2
\partial_-)$ ({\it i.e.} differentiating along the boundary), one obtains
\be
2\partial_+^2 Y+2u^2\partial_-\partial_+ Y =  \lambda A \partial_+u
+\partial_+^2 \ln u \,.
 \label{nggw2}
\ee
In Kruskal gauge the general solution (\ref{xysl}) becomes
\be
Y = -\frac{1}{\gamma}
\left(\lambda^2x^+x^-+x^+P_+-\frac{M}{\lambda}+\frac{\gamma+2}{4}
\ln(-\lambda^2x^+x^-)\right).
\ee
The derivatives of $Y$ are then
\beq
\partial_+Y &=& -\frac{1}{\gamma}
\left(\lambda^2x^-+P_++\frac{\gamma+2}{4x^+}\right) ,
\nonumber\\
\partial^2_+ Y &=& -\frac{1}{\gamma}
\left(T^f_{++}-\frac{\gamma+2}{4x^{+2}}\right),
\nonumber\\
\partial_-\partial_+ Y &=& -\frac{\lambda^2}{\gamma}.
 \label{kgxq2}
\eeq
Substituting these relations into (\ref{nggw2}) leads to
\be
\partial_+^2 \ln u +\lambda A {\partial_+u}
+\frac{2\lambda^2u^2}{\gamma} -\frac{\gamma+2}{2\gamma x^{+2}}
=-\frac{2}{\gamma}T^f_{++}  .
 \label{fqnw}
\ee
An alternate form of this equation is obtained by defining
\be
\omega (\sigma^+)= \lambda\sigma^+ + \ln u  \,.
 \label{zwdf2}
\ee
$\omega$ is a useful variable because, unlike $u$, it is a
constant in the vacuum.
The resulting equation
\be
\omega'' + k(\omega)\lambda\omega'
+\frac{\partial V(\omega)}{\partial \omega}\lambda^2
=-\frac{2}{\gamma}T^f_{++}\,,
 \label{wwqn}
\ee
can be interpreted in terms of a particle
moving in a potential subject to a driving force and a non-linear
damping force,
where $k(\omega)=Ae^\omega -1$,
$\frac{\partial V(\omega)}{\partial \omega}
=\frac{2}{\gamma}e^{2\omega}-Ae^\omega+\frac{\gamma-2}{2\gamma}$
and the primes denote differentiation with
respect to $\sigma^+$. Damping arises because boundary energy can be
dissipated into (or absorbed from) the rest of the spacetime.
Note that the damping becomes negative for
sufficiently negative $\omega$. The potential is
\be
V=\frac{1}{\gamma}e^{2\omega}-Ae^\omega+\frac{\gamma-2}{2\gamma}\omega .
\label{wxqn}
\ee
For ${\gamma}>0$, in the vacuum (\ref{vcmm})
we have $T^f_{++}=0$ and the particle sits
at the local minimum of the potential.  The value $\omega_0$ of $\omega$
at the minimum is
\be
e^{\omega_0}=\frac{\gamma}{4}\left( A +\sqrt{ A^2 -A^2_{min} }\right),
\label{wxqnn}
\ee
(where $A_{min}$ is defined in equation (\ref{amin}) ) in agreement with
(\ref{tmln}). For ${\gamma}<0$, the potential goes to minus infinity for large
negative or positive $\omega$, and there is no local minimum.
Substituting $\omega=\omega_0$ into (\ref{zwdf2}), one finds
\be
\frac{\partial x_B^-}{\partial x^+}
= {u^2}
= \frac{e^{2\omega_0}}{\lambda^2x^{+2}} \,,
 \label{xxwq}
\ee
or
\be
-\lambda^2 x^+x_B^- = e^{2\omega_0} \,.
 \label{xttr}
\ee
The vacuum boundary curve is thus a hyperbola in Kruskal
coordinates, which means it is a straight line located at
$\lambda\sigma=\omega_0$ in sigma coordinates.

\subsection{Vacuum Stability}

It does not appear possible to solve analytically for the
boundary trajectory for a general incoming pulse but one
can easily obtain the leading order in a perturbation
expansion in the strength of the
incoming pulse. Linearizing (\ref{wwqn}) around the vacuum
solution one finds
\be
\gamma {\hat \omega}'' +b\lambda {\hat \omega}'
+(b+2)\lambda^2 {\hat \omega}=-2T^f_{++} \,,
 \label{lnwt}
\ee
where ${\hat \omega} = \omega -\omega_0$ and
\be
b = 2e^{2\omega_0}-\frac{\gamma+2}{2} \,.
 \label{bdfn}
\ee
The general solution to the corresponding homogeneous equation is given by
\be
{\hat \omega}=C_+e^{\alpha_+\sigma^+} + C_-e^{\alpha_-\sigma^+} ,
 \label{dwsl}
\ee
where
\be
\alpha_\pm = \frac{\lambda}{2\gamma}
\left(-b \pm \sqrt{b^2-4\gamma b-8\gamma}\right).
 \label{pdfn}
\ee
Stability of the  vacuum under small perturbations requires that both
solutions in (\ref{dwsl}) be exponentially damped. It is straightforward
to show that this is possible only for $\gamma>0$. Even for $\gamma>0$,
the behavior of the solutions depends on the choice of $A$
(or equivalently $\omega_0$).  In this case, stability requires
$b>0$. This condition can be understood by noting that
$b=\frac{\gamma}{\lambda} \partial_{\sigma}Y(\omega_0)$
when $Y$ is in the vacuum configuration (\ref{vcmm}) and therefore both
$b$ and the slope of $Y$ change sign at a minimum value $Y_{min}$ of $Y$.
In other words, the boundary conditions can only stabilize the system
when the boundary is placed on the physical side of $Y_{min}$.

The condition $b>0$ can also be translated into a restriction on $A$,
which is
\be
A>\frac{2}{\sqrt{\gamma+2}}.
\label{arst}
\ee
For $\gamma>2$, this condition is a consequence of $A>A_{min}$ so
all $\gamma>2$ vacuum-compatible boundary conditons are stable.
When $0<\gamma<2$, this is a new restriction. Apparently (for $\gamma \neq 0$)
$A=0$ is never a stable boundary conditon, and one cannot perturb in $A$.

For $\gamma<0$ ({\it i.e.} $N<24$) the perturbations grow exponentially
for any value of $A$, and we know of no stable boundary conditions.
The case $\gamma=0(N=24)$ will be discussed in a separate publication
\cite{stt}.

While $u$ settles back to its vacuum values after the incident
perturbation is reflected, $x_B^-(x^+)$ undergoes a constant shift.
This can be directly seen from equation (\ref{wtfd2}):
\be
\frac{2}{u} \partial_+Y=\lambda A +\frac{1}{u^2} \partial_+u .
 \label{wtfd}
\ee
The right hand side goes to a constant, but $u$ itself vanishes as
$x^+\rightarrow\infty$. This implies that asymptotically
\be
\lambda^2x_B^-+\bar{P}_+=- \frac{e^{2\omega_0 }}{x^+} , \label{smxm}
\ee
so that $x_B^-$ goes to $-\bar{P}_+/\lambda^2$.
The boundary curve (\ref{smxm}) corresponds to a coordinate
transformation of the original vacuum.

Our boundary conditions were constructed so as to be consistent with
the conformal invariance of the bulk theory, which in particular means
that energy is conserved when the fields are reflected from the
boundary. Conformal invariance at ${\cal I}^+$ (${\cal I}^-$)
ensures that the
change in the Bondi mass equals the outgoing (incoming) energy flux.
However conformal invariance  does not guarantee that the total incoming
and outgoing energies are equal: energy could get stuck on the boundary.
To see that this does not happen for sufficiently small
incoming pulses,
consider the following function defined in Kruskal gauge:
\be\label{massn}
m(x^+,x^-)=\frac{\gamma^2}{\lambda} \partial_+Y\partial_-Y
+\gamma\lambda Y + \frac{(\gamma+2)\lambda}{4}
\left( \ln (\gamma Y)-2\right) +\frac{B \lambda}{Y}\,,
\ee
where the constant $B$ is given by
\be
B= \frac{1}{4} e^{-2\omega_0}b^2 Y_0 -\gamma Y^2_0
- \frac{(\gamma+2)}{4}Y_0
\left( \ln (\gamma Y_0)-2\right).
\label{bval}
\ee
$B$ is chosen so that $m$ vanishes on the boundary if $u$ takes its
vacuum form $e^{\omega_0}/\lambda x^+$, while the constant terms in
(\ref{massn}) have been chosen so that $m$
vanishes asymptotically in the vacuum.
For solutions of the equations of motion, which correspond to matter
energy incident on the vacuum, it is straightforward to establish that
$m(x^+,x^-)$ has the following properties:
\be
\lim_{x^\pm\to \pm\infty}m(x^+,x^-)=m(x^\mp),
\label{lmff}
\ee
where $m(x^\mp)$ is the Bondi mass defined in equation (\ref{mbnd}).
Asymptotically the
function $m(x^+,x^-)$ thus provides an alternate definition of the
Bondi mass.

We will now evaluate $m(x^+,x^-)$ along the boundary curve and verify
that it vanishes before and well after the all the incoming
matter is reflected.  In Kruskal coordinates the boundary conditions
(\ref{ybcg}) and (\ref{xbcg}) imply that along the boundary
$Y=Y_0$ and
\be\label{mbound}
\partial_+Y\partial_-Y = -\frac{1}{4}
\left(\lambda A +\frac{1}{u^2}\partial_+u\right)^2 \,.
\ee
We have already seen that $u$ settles back to its vacuum form after a small
pulse is reflected from the boundary.  In fact, the right hand side
of (\ref{mbound}) goes to the same constant in the two limits
$x^+\to 0$ and $x^+\to \infty$ and the boundary energy thus vanishes
at future timelike infinity. Since the total energy is conserved,
this implies that the incoming and outgoing energy are equal.

The global behavior of the boundary curves will be analyzed in more
detail for the large-$N$ case in Section~4.

\subsection{A Disaster}

We have seen how conformally invariant boundary conditions, imposed at
a boundary placed on the weak coupling side of $Y=0$, ensure that
small pulses incoming from ${\cal I}^-$
are reflected (in a distorted form) up to
${\cal I}^+$.  The behavior for large pulses is quite different.
Consider a pulse which begins (in Kruskal gauge) at an initial
$x^+_i$ and has total momentum $\bar{P}_+$. It was seen in
Section~\ref{secii} that, in the absence of a boundary, the mass
plunges to minus infinity at a point on ${\cal I}^+$,
which is located at $x^-=-\bar{P}_+/\lambda^2$.
This behavior is potentially changed by the presence of the
boundary. The pulse first reaches the boundary at
\be
(x^+,x^-)=\left(x^+_i,-\frac{e^{2\omega_0}}{\lambda^2x^+_i}\right).
\ee
By causality the behavior on ${\cal I}^+$ can not be
affected by boundary reflection prior to
$x^-=-e^{2\omega_0}/\lambda^2x^+_i$. Thus for
\be
\bar{P}_+>e^{2\omega_0}/x^+_i \,,
\ee
the mass still plunges to minus infinity.
{\em No boundary condition at the origin can possibly avert this
disaster for sufficiently large incoming
momentum $\bar{P}_+$.} Since the causal past of the point
$x^-=-\bar{P}_+/\lambda^2$ on ${\cal I}^+$ may
include only regions of weakly coupled dynamics, this disaster
can not in general be averted by modifications of strongly coupled dynamics.
Aversion of this disaster requires fundamentally new input, such as the
inclusion of topology-changing processes. Such considerations are beyond the
scope of the present work, but will be discussed in \cite{asbs}.

It should be noted that this disaster does not necessarily imply a sickness
in the $X,Y$ conformal field theory itself:
rather it arises in the transcription
from the $X,Y$ conformal field theory to a $\rho,\phi$ theory of dilaton
gravity.  The point $\lambda^2x^-=-\bar{P}^+$
is at a finite distance in the fiducial
metric used to regulate the $X,Y$ conformal field theory, and the $X,Y$
fields can be continued past this point. The reflected pulse, (and the
information it carries) eventually comes back out.
However, this occurs ``after the
end of time" as measured by the physical metric $ds^2=-e^{2\rho}dx^+dx^-$.
\newpage

\section{The Large $N$ limit}
\label{largen}

An interesting special case (considered previously by Chung and
Verlinde \cite{verdan}) of our equations is obtained by taking the
limit of a large number of matter fields, {\it i.e.}
$\gamma\rightarrow\infty$.
In taking this limit, $A$, $\lambda^2$, $Y$, $2\tilde X=2X-\ln (N/12)$
and $\tilde T^f_{++}=\frac{12}{N}T^f_{++}$
are held fixed.  The $\tilde X$ and $Y$ OPE's vanish in this limit
and corrections to the semi-classical approximation are systematically
suppressed.

The large $N$ formulae are derived as in the previous
sections and we collect a few of them here.
The semi-classical action (\ref{xctn}) becomes,
\be
S=\frac{N}{12\pi} \int d^2\sigma
\left[-\partial_+\tilde X\partial_-\tilde X
+\partial_+ Y\partial_-Y
+\lambda^2e^{2(\tilde X- Y)}+\frac{6}{N}
\sum^N_{i=1} \partial_+f_i\partial_-f_i\right] \,,
 \label{lnl1}
\ee
and the ++ constraint equation can be written,
\be
 \partial_+Y\partial_+ Y
-\partial_+\tilde X\partial_+\tilde X
+\partial^2_+\tilde X + \tilde T^f_{++} = 0 \,.
 \label{lnl2}
\ee
The solution corresponding to incoming matter (without reflection) is
in sigma gauge,
\be
Y=\tilde X -\lambda\sigma=e^{2\lambda\sigma}
   -\frac{1}{\lambda}e^{\lambda\sigma^+} \tilde P_+(\sigma^+)
+ \frac{1}{\lambda}\tilde M(\sigma^+)
   -\frac{\lambda}{2}\sigma \,,
 \label{lnl3}
\ee
where $\tilde P_+$ and $\tilde M$ are constructed from
$\tilde T^f_{++}$.  The boundary conditions
(\ref{bctw}) are unchanged.  The boundary equation becomes,
\be
\omega''+\lambda(Ae^{\omega}-1)\omega'
+\lambda^2(2e^{2\omega} -Ae^{\omega}+\frac{1}{2} )
=-2 \tilde T^f_{++} \,.
 \label{lnl4}
\ee
In the vacuum
\be
e^{\omega_0}=\frac{1}{4}\left(A+\sqrt{A^2-4}\right).
\label{rfrf}
\ee
For $A<2$ there is no vacuum.
The general solution of the linearized equation is given by
(\ref{dwsl}) with exponents,
\be
\alpha_\pm = \frac{\lambda}{2}\left(
-\tilde b \pm \sqrt{\tilde b^2 - 4\tilde b}\,\right) \,,
 \label{lnl5}
\ee
where $\tilde b = 2e^{2\omega_0}-\frac{1}{2}$.
As before, the boundary curve is stable under small perturbations
provided $A>2$.

We wish to understand the global behavior of solutions to (\ref{lnl4}).
It turns out that the equation (\ref{lnl4}) has a fixed point at
$\omega =-\infty$ as well as at $\omega=\omega_0$. In order to study
this point it is convenient to introduce yet another boundary variable,
\be
v=e^{\omega}=e^{\lambda\sigma^+}u.
 \label{lnss}
\ee
In terms of $v$ the boundary equation is,
\be
v''-\frac{1}{v}(v')^2+\lambda (Av-1)v'
+\lambda^2 (2v^3-Av^2+\frac{v}{2})
=-2v\tilde T^f_{++}.
\ee
The phase portrait (for $\tilde T^f_{++}=0$) is
plotted in figures 1a, 1b and 1c for the overdamped case $A=3.5$ and in
figure 2 for
the underdamped case $A=2.1$ .
The vacuum ($v=e^{\omega_0},~v'=0$) is the
only fully stable fixed point of the equations. However there is
a curious degenerate fixed point at the origin $v=v'=0$,
which is attractive for negative $v'$ and repulsive for positive $v'$.
This unusual behavior arises because of an exact degeneracy
of the equations at $v=0$.
(This degeneracy could be lifted by higher order
corrections to our equation.)
\vbox{{\centerline{\epsfxsize=5.00in \epsfbox[0 0 475 427]{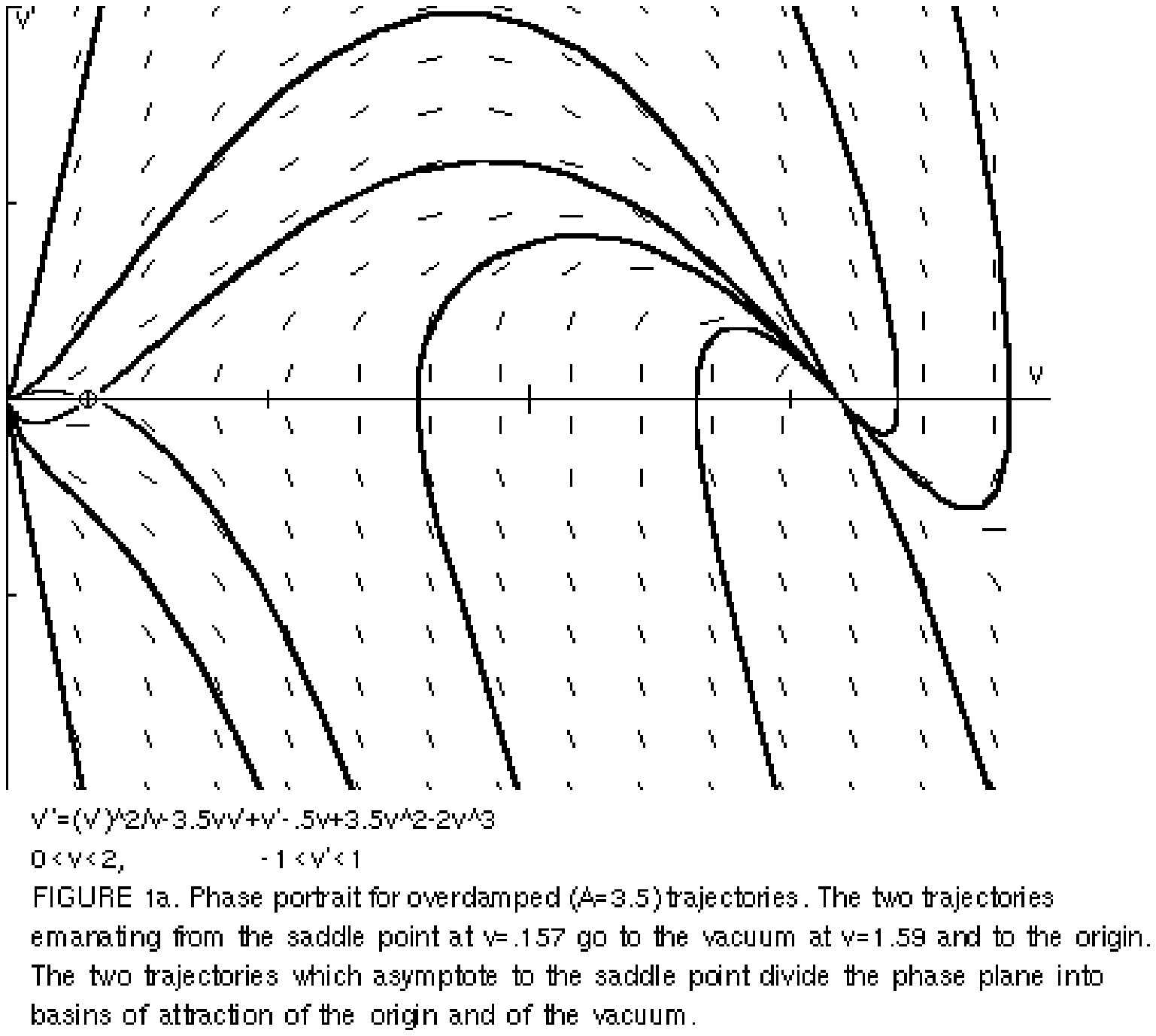}}}}
\vbox{{\centerline{\epsfxsize=5.00in \epsfbox[0 0 475 427]{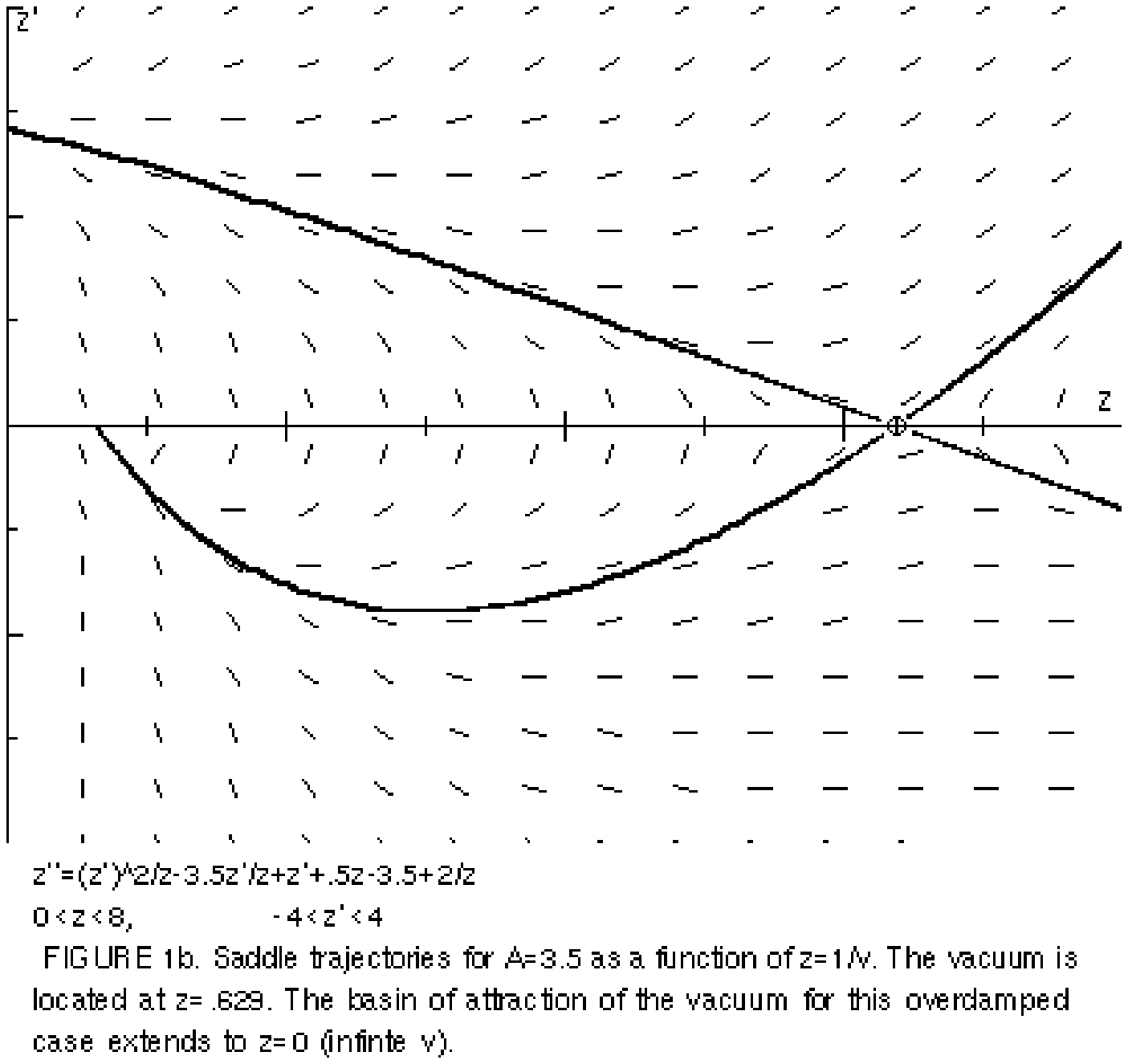}}}}
Near the origin, the equation is dominated by the
terms of degree one in $v$. This observation leads to the general
solution near the origin
\be
v\sim ae^{c e^{\lambda\sigma^+}+\frac{\lambda}{2}\sigma^+},
\label{fcpt}
\ee
where $a>0$ and $c$ are integration constants. A blow-up of the
phase portrait near the origin appears in figure 1c.
Solutions with $c<0$
are approaching the degenerate fixed point,
while those with  $c>0$ are emerging
from it. Emerging trajectories have a universal
initial slope of $\frac{\lambda}{2}$,
while approaching trajectories have an asymptotically
infinite ``vertical'' slope.
\vbox{{\centerline{\epsfxsize=5.00in \epsfbox[0 0 475 427]{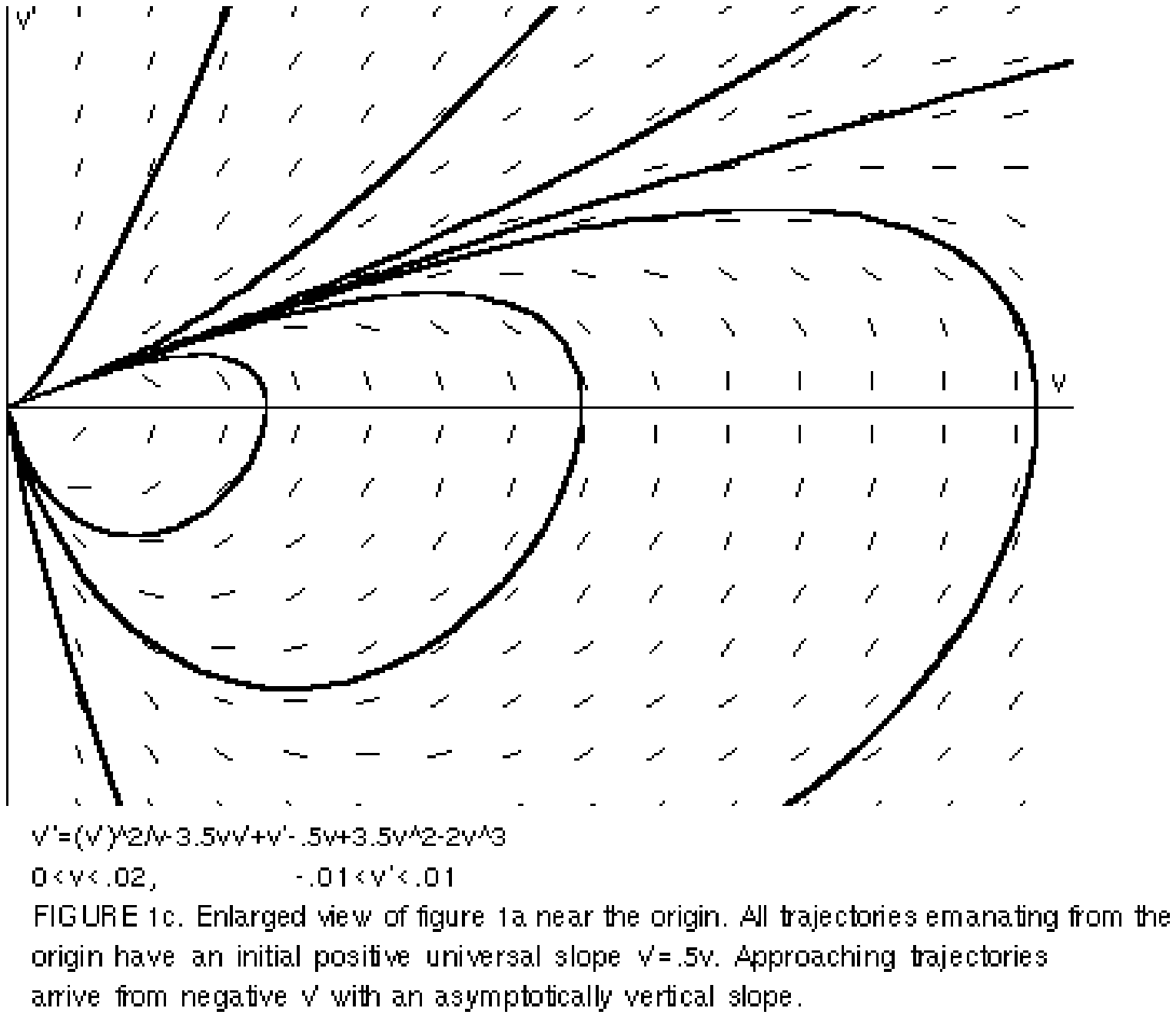}}}}
It is instructive to
estimate the boundary mass along these approaching trajectories.
At large $N$, the mass formula (\ref{massn})
becomes
\be\label{massf}
m(x^+,x^-)=\frac{1}{\lambda} \partial_+Y\partial_-Y
+\lambda Y + \frac{\lambda}{4}
\left( \ln Y-2\right) +\frac{B\lambda}{Y}\,,
\ee
where the constant $B$ is
adjusted as in (\ref{bval}) so that $m$ vanishes on the boundary in the vacuum.
The formula (\ref{mbound}) which gives the variation of the mass
along the boundary is unchanged.  One finds for $c<0$ that on the boundary
\be\label{minf}
m \sim -\frac{c^2\lambda}{4a^2}
e^{-2ce^{\lambda\sigma^+}+{\lambda}\sigma^+}\, ,
\ee
so that the mass plummets to minus infinity along approaching ($c<0$)
trajectories.

 If a shock wave of total mass $M$ is sent at the boundary
along $\sigma^+=0$,
the initial data just after the shock wave is
$v(0)=e^{\omega_0}, v'(0)=-2e^{\omega_0}M$, where
$e^{\omega_0}$ is the vacuum value of $v$.
As can be seen from figures 1b or 2, for sufficiently large $M$, the initial
data lie in the basin of attraction of the degenerate fixed point.

The pathological behavior along trajectories heading to the
degenerate fixed point is directly associated to black hole formation,
as follows. By definition, a black hole is a region in which
$\partial_+ Y<0$ \cite{revs}.
The boundary may at some point enter such a region,
and thereafter be in or behind a black hole.  When $\partial_+ Y$
crosses zero a spacelike
line along which $Y=Y_0$ will branch off from the timelike boundary line
along which $Y=Y_0$. At the branching point, both the tangential and
normal derivatives of $Y$ along the boundary will vanish.
The condition for this to occur follows directly
from the boundary condition (\ref{wtfd2}),
and in terms of the variable $v$ it reads,
\be
v'_{cr}(v)=\lambda (v-Av^2).
\label{vcrt}
\ee
For the special
case of shock waves this reduces to the energy threshold
$M=2\lambda(e^{2\omega_0}-\frac{1}{4})$ for an apparent horizon to form
before the shock wave reaches the boundary.
It is easy to see
\vbox{{\centerline{\epsfxsize=5.00in \epsfbox[0 0 475 427]{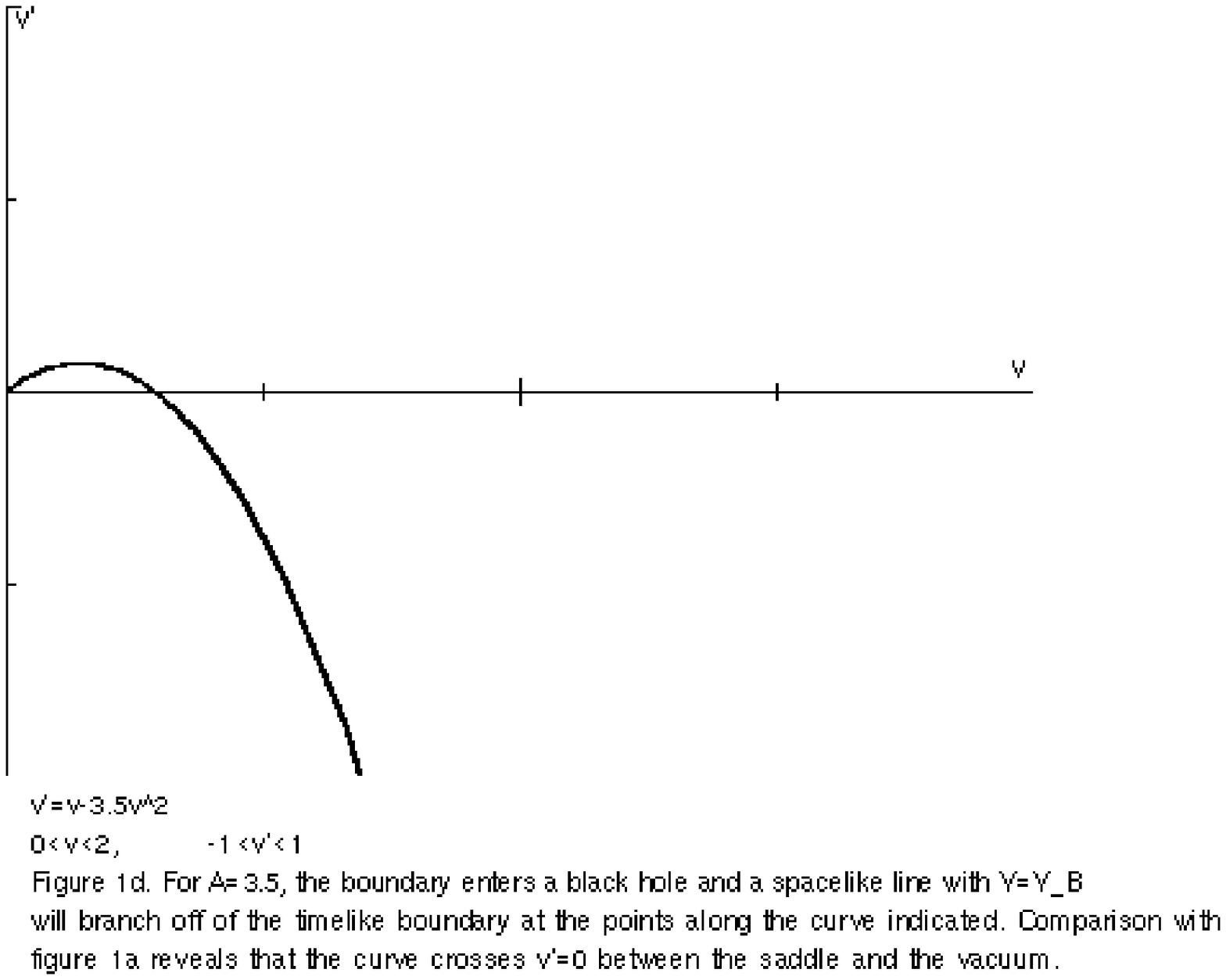}}}}
from figure 1d that any curve which begins at the vacuum
($v(0)=e^{\omega_0}, v'(0)=0$) and asymptotes to the
degenerate fixed point at the origin
from below necessarily crosses the line
$v'=v'_{cr}$. (The converse is not true, as may be seen
from the figures: the boundary may
enter and leave a region where $\partial_+ Y<0$ and
still settle back to the vacuum.)
Thus the runaway behavior lies behind a black hole, and is another form
of the ``disaster'' discussed at the end of Section~\ref{seciii}.

\vbox{{\centerline{\epsfxsize=5.00in \epsfbox[0 0 475 427]{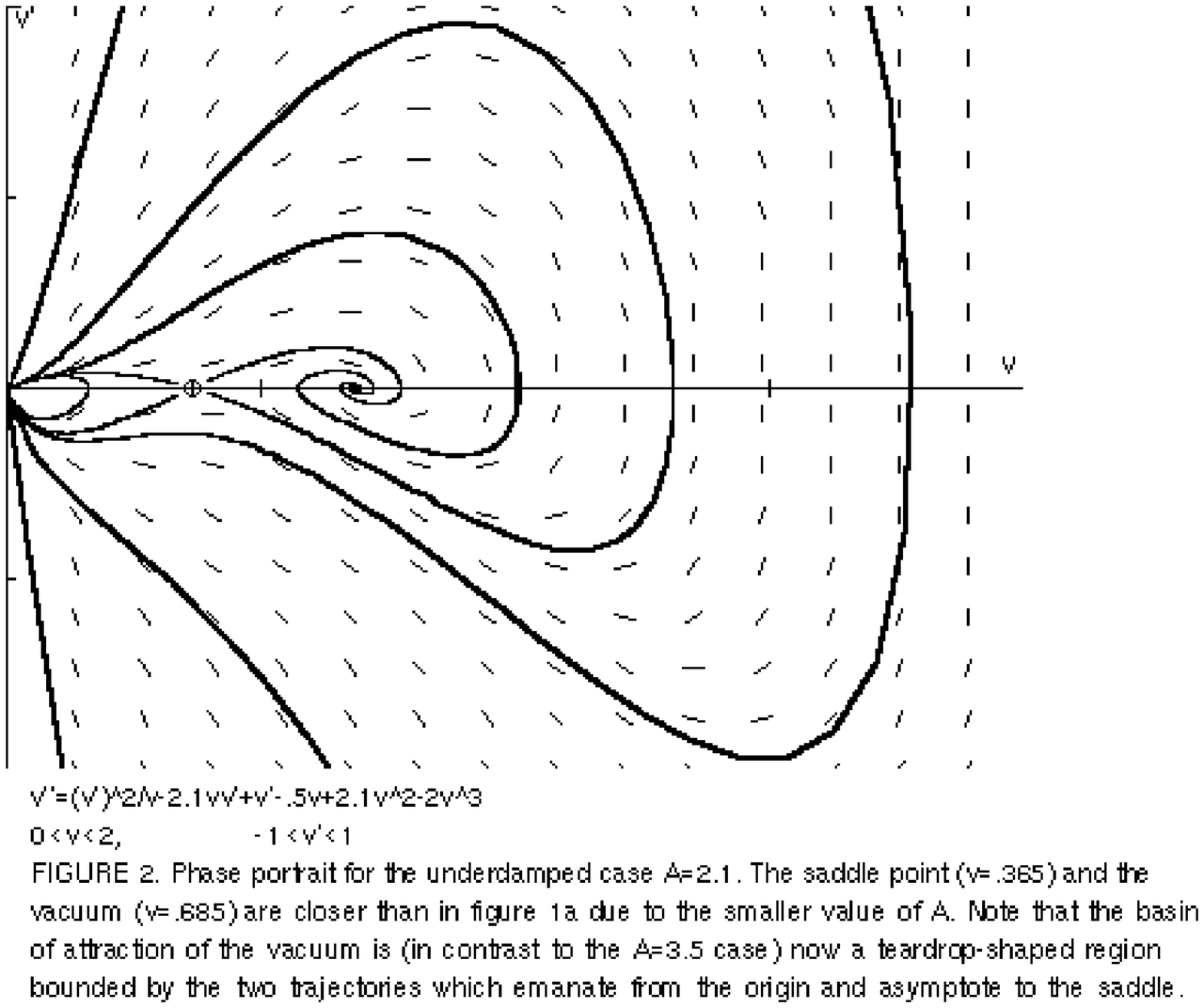}}}}

\section{The $\lambda^2=0$ Limit}
\label{seciv}

In this section we consider the theory obtained by setting $\lambda^2=0$
in the action (\ref{xctn}). Although black holes are absent in this limit,
it is of special interest for several reasons.
First, the theory remains semi-classically soluble
even after the non-linear boundary conditions (\ref{bctw}) are imposed.
The resulting theory is comparable in complexity to Liouville theory,
and is of interest in its own right as a non-trivial conformal field
theory with boundary interactions. It is similar to a boundary theory
which was recently exactly solved in \cite{cklm,jplt} and analogous
methods may be applicable here.
Second, it arises as an effective short distance theory
for (\ref{xctn}) at scales short relative to $\lambda^{-1}$. Finally,
it bears the following direct relationship to the $\lambda^2\neq0$ theory.
The fields $X$ and $Y$ for nonzero $\lambda^2$ can be expressed in terms
of free fields $x$ and $y$,
\be
\partial_+\partial_- x = \partial_+\partial_-y = 0,
 \label{ffxy}
\ee
via the relation
\beq
X - Y &=& x- y, \nonumber \\
Y &=& y + \frac{\lambda^2}{\gamma}
\int d\sigma^+\int d\sigma^-e^{2(x-y)} \,.
 \label{frdf}
\eeq
The gravitational stress tensor (\ref{xyst}) becomes simply
\be
T^g_{++} =
\gamma \left(
 \partial_+y\partial_+y
-\partial_+x\partial_+x
+\partial^2_+ x
\right) \,.
 \label{xrst}
\ee
The OPE's
\beq
\partial_+ x(\sigma^+)\partial_+x({\sigma^+}') &=&
\frac{ -1}{4\gamma(\sigma^+-{\sigma^+}')^2} \,, \nonumber\\
\partial_+ y(\sigma^+)\partial_+y({\sigma^+}') &=&
\frac{1}{4\gamma(\sigma^+-{\sigma^+}')^2} \,,
 \label{rxpn}
\eeq
imply the original OPE's for $X$ and $Y$
when substituted in to (\ref{frdf}). Thus the field
redefinition (\ref{frdf}) transforms the bulk $\lambda^2\neq0$ theory
into the manifestly free $\lambda^2=0$ theory.

Now let us apply the boundary condition (\ref{bctw}) for $Y_0=0$ directly
on the free fields $x=x_++x_-$ and $y=y_++y_-$,
\beq
\partial_+x -\partial_-x &=& \alpha e^x, \nonumber\\
 y &=& 0 \,, \label{bctr}
\eeq
where $\alpha=\lambda A$.
Note that this is {\it not\/} the same as rewriting (\ref{bctw}) in terms
of the free fields $x$ and $y$, in which case we would have a rather
complex set of boundary conditions for the free fields.
It is straightforward to solve (\ref{bctr}) for the outgoing fields
$x_-$, $y_-$ in terms of the incoming fields $x_+$, $y_+$.
The result is
\beq
y_-(\tau) &=& - y_+(\tau) , \nonumber\\
x_-(\tau) &=& x_+(\tau)
-\ln \left(\alpha \int^\tau d\tilde\tau\,e^{2x_+(\tilde\tau)}\right) .
 \label{rbcn}
\eeq
A convenient gauge is
\be
x_+(\omega^+)=\omega^+ ,
 \label{xpwp}
\ee
and (\ref{rbcn}) then implies
\be
x_-(\omega^-) =-\omega^--\ln\frac{\alpha}{2} .
 \label{xmwm}
\ee
The constraints further imply that
\be
\partial_+y\partial_+y=1-\frac{1}{\gamma}T^f_{++} \,,
 \label{yygt}
\ee
or
\be
y_+(\omega^+) =
\int^{\omega^+} d\tilde\omega^+ \,
\sqrt{1-\frac{1}{\gamma}T^f_{++}(\tilde\omega^+)} .
 \label{ypwp}
\ee
The reflection condition (\ref{rbcn}) gives
\be
y_-(\tau)=-y_+(\tau).
 \label{ymyp}
\ee
Evidently the configuration
\be
x_\pm = y_\pm =\pm \omega^\pm \,,
\label{xypm}
\ee
is stable under small matter perturbations from ${\cal I}^-$.
Unfortunately (\ref{xypm}) does not quite correspond to the $x,y$
configuration of the vacuum (\ref{vcmm}) of the $\lambda\neq0$ theory.
The $x,y$ configuration which does correspond to (\ref{vcmm}) is
apparently not stable under perturbations, and thus (\ref{rbcn})
does not directly translate into stable boundary conditions for the
$\lambda\neq0$ theory.

\section{Acknowledgements}

We are grateful to B. Birnir, S.~Giddings, J.~Polchinski, and especially
S.~Trivedi for useful discussions. The figures were plotted on a Macintosh
computer with McMath. This work was supported in part by
DOE grant DOE-91ER40618 and NSF~Grant~No. PHY89-04035.

\end{document}